\documentclass{ws-procs975x65}
\usepackage{graphicx,units,color}
\usepackage{epstopdf}
\begin{document}

\title{Very high energy gamma-rays and the Hubble parameter}

\author{Alexia Gorecki, Aur\'{e}lien Barrau$^*$}

\address{Laboratoire de Physique Subatomique et de Cosmologie, CNRS/UJF/INPG\\
53, rue des Martyrs 38026 Grenoble Cedex, France\\
$^*$E-mail: barrau@lpsc.in2p3.fr}

\author{Julien Grain}
\address{Institut d'Astrophysique Spatiale, Universit\'{e} Paris-Sud 11, CNRS\\
B\^{a}timents 120-121, 91405 Orsay Cedex, France}

\author{Elisabetta Memola}
\address{Istituto Nazionale di Fisica Nucleare, INFN - Milano Bicocca \\
Piazza della Scienza 3,
20126 Milano, Italy}

\begin{abstract}
A new method, based on the absorption of very high-energy gamma-rays by the cosmic infrared
background, is proposed to constrain the value of the Hubble constant. As this value is
both fundamental for cosmology and still not very well measured, it is worth developing
such alternative methods. Our lower limit at the 68$\%$ confidence level is
$H_0 >\unit[74]{km/s/Mpc}$,
leading, when combined with the HST results, to $H_0 \approx \unit[76]{km/s/Mpc}$.
Interestingly, this value, which is significantly higher than the usually considered one,
is in exact agreement with other independent approaches based on baryonic acoustic
oscillations and X-ray measurements. Forthcoming data from the experiments HESS-2 and CTA
should help improving those results. Finally, we briefly mention a plausible correlation
between absorption by the extragalactic background light and the absence of observation of
gamma-ray bursts (GRBs) at very high energies.
\end{abstract}

\bigskip

As the very high energy photons emitted by Active Galactic Nuclei (AGN) travel, they
interact with the Cosmic Infrared Background (CIB) photons by electron-positron pair
production. The cross section is maximum for the highest energies
(typically around $\unit[10-20]{TeV}$) and the interaction takes place with
CIB photons in the range $\unit[1-100]{\mu m}$.
Assuming a well defined CIB energy density, the observed AGN
spectrum can be unfolded. The absorption factor entering the unfolding procedure
depends on the Hubble parameter, therefore it can be easily constrained
by comparing the unfolded spectrum with theoretically ``allowed" spectra.
Quite simply, the smaller the Hubble constant, the higher the distance to the source, the
higher the absorption, the harder the unfolded  spectrum. As a consequence,
a too small value of the Hubble parameter can eventually lead to an unphysical unfolded
spectrum.

The CIB is the relic emission of the formation and evolution of galaxies of all types and
star-forming systems.
The near-IR CIB, from $\unit[1~\textrm{to}~10]{\mu m}$, arises mainly from
the stellar component of galaxies (probing the evolution at early times).
For higher wavelengths, between
$\unit[10~\textrm{and}~200]{\mu m}$, the radiation is mostly due to dusty Ultra Luminous
InfraRed Galaxies. The measure of the CIB spectrum is challenging as it is difficult to
distinguish between the foreground, the atmosphere, and the instrument emission itself.  
Therefore, different methods are used in order to determine the CIB all along the
emission range. We have considered an
up-to-date CIB measurement inventory\cite{bernstein}$^-$\cite{wright}\!,
which satisfactorily constrains the spectrum.

To turn down the unfolded spectrum into a lower limit on $H_0$, one has to assume a class
of ``possible" source emission spectra. It is well known that the
Inverse-Compton (IC) upper energy range bump mimics the concave Synchrotron bump
(observed at
lower energy). Moreover, there is no physical process to inject energy above the IC bump,
and the Klein-Nishina effect tends to soften the spectrum at high energy. As a consequence, the
main hypothesis of this study is to assume that the AGN spectrum must be concave at high
energy, which is in agreement with the approach of Ref.~\refcite{renault} and Ref.~\refcite{guy}.
The blazar Mrk 501, with redshift $z = 0.034$,
is a good source candidate because it
has been intensively observed in a broad dynamical range between $\unit[400]{GeV}$ and
$\unit[21]{TeV}$ during the 1997 flare by the experiments CAT
and HEGRA,
as reported in Ref.~\refcite{aharonian} and Ref.~\refcite{guy}.

\begin{figure}
\begin{center}
\includegraphics[scale=0.5]{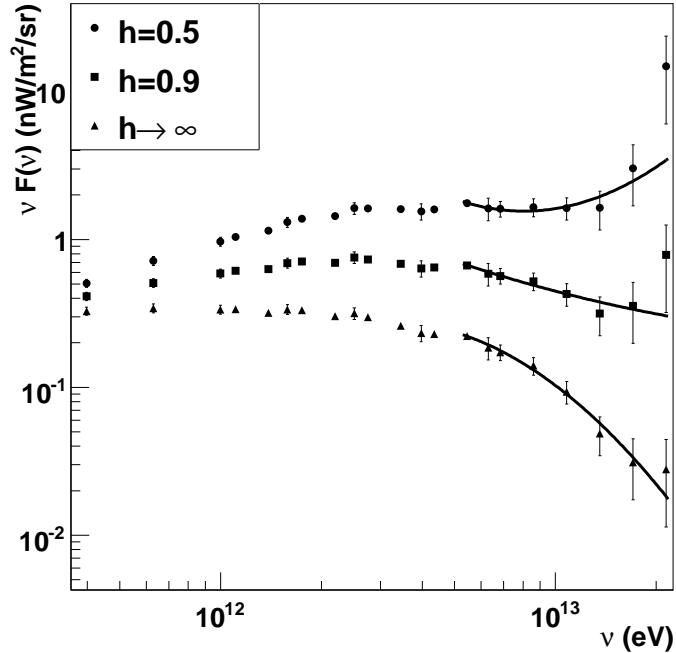}
\caption{Unfolded gamma-ray spectra and associated fits for (from bottom to top): no CIB
(equivalent to $H_0\rightarrow \infty$), $H_0=\unit[50]{km/s/Mpc}$, {\rm and} $H_0=\unit[90]{km/s/Mpc}$.\label{AGN_spectrum}}
\end{center}
\end{figure}

The unfolded spectrum is related with the observed spectrum by a
multiplicative factor which is the exponential of the optical depth $\tau$.
This optical
depth depends on the cross section $\sigma(E_{AGN},\epsilon_{CIB})$
for $\gamma-\gamma$ pair production\cite{heitler}
being $E$ and $\epsilon$ the observed energies at $z$=0,
on the CIB energy density $n(\epsilon_{CIB})$, and on the Hubble parameter as follows:
$\tau\propto\frac{1}{H_0}\int n(\epsilon_{CIB})\sigma(E_{AGN},\epsilon_{CIB})d\epsilon_{CIB}$.
A Monte Carlo simulation was performed: at each step, for each wavelength, the CIB
flux is randomly selected according to a Gaussian law centered at the best CIB estimate at
the considered wavelength (the width is taken as the experimental error bar). This
procedure is repeated for
several $H_0$ values. Figure \ref{AGN_spectrum} shows the unfolded spectrum
$\nu F_{unfolded}(\nu)=e^{\tau(H_0)}\nu F_{obs}(\nu)$, for different values of $H_0$. One
can see that if the value of $H_0$ is too small, the unfolded spectrum is no longer
concave, which is unphysical. Taking into account the uncertainties, the concavity
requirement leads to $H_0 >\unit[74]{km/s/Mpc}$ at the $68\%$ confidence level (and to
$H_0 >\unit[54]{km/s/Mpc}$ at $90\%$ confidence level)\cite{barrau}\!.

Finally, it is worth mentioning that absorption by
the extragalactic background light  could
very well play an important role in understanding the relatively
rare population of GRBs observed above 10 GeV.
So far, the highest energy emitted from a gamma-ray burst
has been measured by the {\em Fermi} Large Area Telescope, that
observed a photon\cite{abdo} of 33.4$^{+2.7} _{-3.5}$\,GeV
from the GRB 090902B at z=1.822.
Although slightly smaller than Stecker values\cite{stecker}\!, our raw estimates of the
optical depth is of the same order ($\tau\sim 5$). This would lead to huge absorption factors of
the order of 150, making the unfolded spectrum quite unphysical. Evolution effects should of course be
correctly taken into account, but the situation is quite puzzling.

\end{document}